\documentclass[letterpaper,12pt]{article}
\usepackage{amsmath}
\usepackage{amsfonts}
\usepackage{amssymb}
\usepackage{mathrsfs}
\usepackage{multicol}
\usepackage{color}
\usepackage{graphicx}
\usepackage{pstricks}
\usepackage{pst-node}
\usepackage{wrapfig}
\usepackage{enumerate}
\usepackage{caption}
\usepackage{cite}
\usepackage{subfigure}
\usepackage{amsfonts,amssymb}
\usepackage{amsthm}
\usepackage{amscd}
\definecolor{gray1}{gray}{0.1}
\definecolor{gray2}{gray}{0.2}
\definecolor{gray3}{gray}{0.3}
\definecolor{gray4}{gray}{0.4}
\definecolor{gray5}{gray}{0.5}
\definecolor{gray6}{gray}{0.6}
\definecolor{gray7}{gray}{0.7}
\definecolor{gray8}{gray}{0.8}
\definecolor{gray9}{gray}{0.9}

\newpsobject{showgrid}{psgrid}{subgriddiv=1,griddots=10,gridlabels=6pt}

\definecolor{dark-green}{rgb}{0,0.7,0}
\definecolor{dark-blue}{rgb}{0,0.2,0.5}
\definecolor{med-blue}{rgb}{0,0.7,1}
\definecolor{mblue}{rgb}{0,0.2,1}
\definecolor{cnc}{rgb}{0.8,0,0}
\definecolor{light-red}{rgb}{1,0.8,0.8}
\definecolor{dark-yelow}{rgb}{1,0.8,0}
\definecolor{light-blue}{rgb}{0.8,0.9,1}
\definecolor{verylight-blue}{rgb}{0.93,0.95,1}
\definecolor{light-yelow}{rgb}{1,0.9,0.8}
\definecolor{grey}{gray}{0.88}

\newpsobject{showgrid}{psgrid}{subgriddiv=1,griddots=10,gridlabels=6pt}

\begin{document}

\thispagestyle{empty}

\setlength{\abovecaptionskip}{10pt}

\begin{center}
{\Large\bfseries\sffamily{Inflation inside non-topological defects 
and scalar black holes}}
\end{center}
\vskip 1mm

\begin{center}

\small \textbf{Yves Brihaye $^{(1)}$, Felipe C\^onsole $^{(2)}$ and  Betti Hartmann $^{(2), (3), (4)}$} \\ [0.5cm] 
\small $^{(1)}$ Universit\'e de Mons, Place du Parc, 7000 Mons, Belgium\\
$^{(2)}$ Instituto de F\'isica de S\~ao Carlos, Universidade de S\~ao Paulo, S\~ao Carlos, S\~ao Paulo 13560-970, Brazil\\
$^{(3)}$ Institut f\"ur Physik, Carl-von-Ossietzky Universit\"at Oldenburg, 26111 Oldenburg, Germany\\
$^{(4)}$  Department of Physics and Earth Sciences, Jacobs University Bremen, 28759 Bremen, Germany
\end{center}
\vspace{1.5cm}
\begin{abstract}
In this paper, we demonstrate that a phenomenon described as {\it topological inflation} during which inflation
occurs inside the core of topological defects, has a non--topological counterpart. This appears in a simple set-up
containing Einstein gravity coupled minimally to an electromagnetic field as well as a self-interacting, complex
valued scalar field. The U(1) symmetry of the model is unbroken and leads to the existence of
globally regular solutions, so-called boson stars, that develop a horizon for sufficiently strong gravitational coupling.
We also find that the same phenomenon exists for
black holes with scalar hair. \\
\vspace{0.1cm}\\
{\it Manuscript prepared for special edition of ``Symmetry'' in celebration of Yves Brihaye's career.}

\end{abstract}
\vspace{0.2cm}
 

\section{Introduction}

The theory of General Relativity (GR) is the best tested theory of gravity developed up to date. The classical tests, within the solar system, where the deviations from Newtonian gravity are small, helped to promote GR to one of the fundamental theories of nature. For strong gravitational fields, i.e. when the deviations from Newtonian gravity become important, observations and tests have now become available. An excellent laboratory to consider strong gravitational field effects are black holes (BHs) and neutron stars (NSs). It’s a common belief that BHs are ``simpler''to describe than NSs partly due to the lack of knowledge of the equation of state of the matter making up the NS. BHs, in fact, are still believed to follow the so-called no-hair conjecture \cite{Ruffini:1971bza}, which states that all stationary asymptotically flat BHs are fully characterized by global charges associated with a Gauss law. There are numerous counter-examples for the conjecture when considering non-linear matter fields. But the situation for scalar fields is very different. A number of no-scalar-hair theorems were put forward (see \cite{herdeiro2018asymptotically} for a recent review) and scalar fields, apparently, should be trivial around a stationary and asymptotically flat BH spacetime. 

One process where a scalar field can become non-trivial in a BH spacetime is through spontaneous scalarization. It was first discussed in a scalar-tensor theory of gravity around a NS \cite{Damour:1993hw}, in which the scalar field couples to the trace of the energy-momentum tensor and can obtain non-vanishing values even if its asymptotic value is zero.. The corresponding phenomenon has gained considerable attention in black hole physics \cite{Silva_2018}.

The view on no-hair theorems for minimally coupled scalar fields has changed since the discovery of hairy Kerr BHs in a model where a massive complex scalar field is minimally coupled to gravity \cite{Hod:2012px, Herdeiro_2014}. To circumvent the no-scalar-hair theorems, it’s necessary to assume a harmonic dependence on the time and azimuth coordinates. Then the so-called synchronization condition $\omega/m = \Omega_H$ must be imposed, where $\omega$ is the scalar field frequency, $m$ an integer and $\Omega_H$ the horizon angular velocity. The configuration of the test scalar field outside the event horizon was dubbed ``scalar cloud" and when taking backreaction into account leads to the existence of hairy Kerr BHs solutions. As the horizon radius approaches zero, the solution is reduced to a spinning \emph{boson star}. 

Boson stars (BSs) are regular, stationary, and localized solutions to the Einstein-Klein-Gordon system of equations, formed by a complex scalar field with a continuous U(1) symmetry, which gives rise to a globally conserved Noether charge. They are the self-gravitating counterparts
of $Q$-balls \cite{coleman}.

Their size can range from the atomic scale up to the size of supermassive BHs, depending on the choice of the scalar potential (see \cite{Schunck_2003} for a review), and can be used as models for dark matter particles \cite{Eby_2016} and BHs mimickers \cite{Guzm_n_2009}, for example. The absence of an event horizon, however, can lead to significant changes in the propagation of light rays when compared to the spacetime of a BH \cite{Macedo_2013}. Charged BSs were studied in \cite{Brihaye:2014gua} for a self-interacting scalar potential whose motivation comes from supersymmetric extensions of the standard model and originally, uncharged $Q$-balls had been discussed \cite{susy1,susy2,susy3}. A combination of the attractive effects of gravity and the repulsive effects of electromagnetism can render the charged BS stable.

Further studies concerning scalar clouds were considered subsequently for Reissner-Nordstr\"om (RN) spacetime \cite{Herdeiro:2020xmb}, where for a non-trivial configuration of the gauged scalar field it was shown that it is necessary to add self-interactions in the scalar potential and the resonance condition to be satisfied, $\omega = q V(r_h)$, where $q$ is the scalar coupling constant and $V(r_h)$ the electric potential on the horizon.

Gauged scalar clouds in the Schwarzschild BH were considered in \cite{Herdeiro:2020xmb} and \cite{brihaye_hartmann2020}. In this case, the background is fixed by the Schwarzschild metric and the differential equations for the electric potential and the scalar field are coupled. It was shown that the scalar clouds exist for some range in the gauge coupling and it was also found numerically in \cite{brihaye_hartmann2020} that two different solutions exist for the same values of the gauge coupling constant and the electric potential at infinity. When backreaction is taken into account, the solutions exist up to a maximal value of the gravitational constant and lead to two distinct situations: (i) an extremal BH with a diverging derivative of the scalar field at the horizon and (ii) a RN-de Sitter solution with a screened electric charge. 

In this paper, we extend the results of \cite{brihaye_hartmann2020} to include globally regular space-times with the same matter field content.
The corresponding solutions are charged $Q$-balls (in a Minkowski space-time) and boson stars. IN the following, we will
demonstrate that the phenomenon described above does neither depend on the details of the scalar self-interaction nor 
on the fact that the space-time possesses {\it a priori} a horizon. 
Our paper is organized as follows~: in Section 2, we discuss the model and equations of motion, while Section 3 contains our results
on black hole and globally regular space-times, respectively. We end with a discussion in Section 4. 
 
\section{Model}
Here we will discuss solutions to the following (3+1)-dimensional model~:
\begin{equation}
\label{eq:action}
{\cal S}=\int {\rm d}^4 x \sqrt{-g} {\cal L} \ \ \ , \ \ \ 
    {\cal L} = \frac{{\cal R}}{16 \pi G} - \left(D_{\mu} \Psi\right)^* D^{\mu} \Psi - \frac{1}{4}F_{\mu \nu} F^{\mu \nu} - U(\vert\Psi\vert) \ .
\end{equation}
This is the Einstein-Hilbert action with ${\cal R}$ the Ricci scalar and $G$ Newton's constant
minimally coupled to a complex valued scalar field $\Psi$ that is charged under a U(1) gauge
field $A_{\mu}$.  $D_{\mu} = \partial_{\mu} - i q A_{\mu}$ is the covariant derivative of the scalar field and $F_{\mu\nu}=\partial_{\mu} A_{\nu} - \partial_{\nu} A_{\mu}$ the field strength tensor
of the U(1) gauge field. The scalar field potential $U(\vert\Psi\vert)$ will turn out to be a crucial
ingredient in our construction in the following. We will choose the potential as follow \cite{susy1,susy2}~:
\begin{equation}
\label{eq:potential}
U(\vert\Psi\vert) = \mu^2 \eta^2 \left[1 - \exp\left(-\frac{\vert\Psi\vert^2}{\eta^2}\right)\right] \ ,
\end{equation}
where $\mu$ corresponds to the mass of the scalar field and $\eta$ is an energy scale.

In the following, we are interested in spherically symmetric and stationary solutions. The Ansatz for the metric is~:
\begin{equation}
     ds^2 = - N(r) (\sigma(r))^2 {\rm d}t^2 + \frac{1}{N(r)} {\rm d}r^2 + r^2({\rm d} \theta^2 + \sin^2 \theta {\rm d} \varphi^2) \ \ , \ \  N(r)=1-\frac{2 m(r)}{r} \ , 
\end{equation}
while the matter fields are chosen according to~:
\begin{equation}
      \Psi(r,t)  = \eta e^{i \tilde{\omega} t} \psi(r) \ \ , \ \ A_0 = \eta V(r) \ \  
\end{equation}      
with $\tilde{\omega}$ a real constant. 
Note that, although the scalar field is time-dependent, the associated energy-momentum tensor
is static and hence likewise the space-time. Defining the following dimensionless quantities
\begin{equation}
\label{eq:rescalings}
x=\mu r \ \ , \ \  \omega=\frac{\tilde{\omega}}{\mu} \ \ , \ \   e=\frac{\eta q}{\mu}  \ \ , \ \  \alpha=4\pi G \eta^2
\end{equation} 
the equations resulting from the variation of the action (\ref{eq:action}) depend only on $e$ and
$\alpha$ and read (with the prime denoting derivative with respect to $x$)~:    
 \begin{eqnarray}
    m' &=& \alpha x^2 \biggl[ \frac{V'^2}{2 \sigma^2} + N \psi'^2 + U(\psi) + \frac{(\omega-e V)^2}{N \sigma^2}\psi^2 \biggr]\\
		\sigma' &=&  2 \alpha x \sigma \biggl[ \psi'^2 + \frac{(\omega - e V)^2}{N^2 \sigma^2}\psi^2 \biggr]
\label{eq:metric}
\end{eqnarray}
for the metric functions and 
 \begin{eqnarray}
      V'' &+&  \left(\frac{2}{x} -\frac{\sigma'}{\sigma}\right) V' + \frac{2(\omega-e V)  \psi^2}{N} = 0 
			\label{eq:eq_v} \\
	\psi'' &+& \left(\frac{2}{x} + \frac{N'}{N} +\frac{\sigma'}{\sigma}\right) \psi' + \frac{(\omega -e V)^2 \psi}{N^2 \sigma^2} - \frac{1}{2N}
			\frac{dU}{d \psi} = 0 \ .
\label{eq:eq_psi}
\end{eqnarray}
for the matter fields. As is obvious, these equations depend only on the combination $\omega - e V(x)$. 
Note that the Lagrangian density and with it the energy density are now given in units of $\eta^2 \mu^2$. 

The asymptotic behaviour of the metric and matter field functions is~:
\begin{equation}
\label{eq:infty_metric}
N(x \gg 1)=1-\frac{2M}{x} + \frac{\alpha Q^2}{x^2} + ..... \ \ , \ \  \sigma(x\gg 1) = 1 + {\cal O}(x^{-4})  \ \ , 
\end{equation}
\begin{equation}
\label{eq:infty_matter}
V(x)=V_{\infty}  - \frac{Q}{x} + ....  \  \ , \  \   \psi(x\rightarrow \infty)\sim \frac{\exp(-\mu_{\rm eff,\infty} x)}{x} + .... \ \ , \  \
\mu_{{\rm eff},\infty}=\sqrt{1 - \Omega^2} \ ,
\end{equation}
where $\Omega^2:=(\omega-e V_{\infty})^2$.
$M$ and $Q$ denote the (dimensionless) mass and electric charge of the solution, respectively. $V_{\infty}$ can be understood
to be a ``chemical potential'', i.e. the resistance of the system against the  additon of extra charges $e$ to the system. 
Note that the scalar field possesses an effective mass $\mu_{\rm eff}$ smaller than the bare mass of the scalar field, which  -- in our dimensionless units -- is equal to unity. 

Next to the electric charge, the solutions possess a Noether charge which results from the unbroken U(1) symmetry of the system. This reads~:
\begin{equation}
\label{eq:noether}
Q_N=\int {\rm d} x \ \frac{2x^2 e v\psi^2}{N\sigma}  \ .
\end{equation}
This quantity can be interpreted as the number of scalar bosons that make up the solution.
For globally regular solutions $eQ_N\equiv Q$, while for black holes, $Q=eQ_N - E_x(x_h) x_h^2/\sigma(x_h)$, where $E_x(x)=-V'(x)$ is the radial electric field of the solution. The second term in this equality is the horizon electric charge and is the consequences of the fact that the horizon corresponds to a surface on which boundary conditions have to be imposed. 
Finally, the temperature $T_{\rm H}$ and entropy ${\cal S}$ of the black hole solutions is given by 
\begin{equation}
 T_H=(4\pi)^{-1}\sigma(x_h) N'\vert_{x=x_h}   \ \ , \ \   {\cal S} = \pi x_h^2 \ . 
\end{equation}

The energy-density $\epsilon$, radial pressure $p_r$ and tangential pressure $p_{\theta}=p_{\varphi}$, respectively,  read~:
\begin{align}
\label{eq:density_pressure}
\epsilon=-T_t^t = \epsilon_1 + \epsilon_2 + \epsilon_3 +  U(\psi)  \\
p_r=T_r^r =  - \epsilon_1 + \epsilon_2 + \epsilon_3 - U(\psi) \\
p_{\theta}=T_{\theta}^{\theta}=\epsilon_1 - \epsilon_2 +\epsilon_3 - U(\psi)   \ .
\end{align}
with
\begin{equation}
\label{eq:EMcomponents}
\epsilon_1=\frac{V'^2}{2\sigma^2}  \ \ , \ \  \epsilon_2= N \psi'^2  \ \ , \ \    \epsilon_3= \frac{(\omega - e V)^2 \psi^2}{N \sigma^2}  \ .
\end{equation}

\section{Results}
In the following, we will discuss the different type of solutions to the equations of motion
(\ref{eq:metric}), (\ref{eq:eq_v}) and (\ref{eq:eq_psi}).  We will start with the discussion
of black holes, for which we have integrated the equations between $x=x_h$, i.e. the horizon and infinity, respectively. 
The solutions correspond to Schwarzschild black holes endowed with charged scalar clouds, so-called $Q$-clouds, when neglecting the backreaction
of the cloud on the space-time, otherwise to charged black holes with scalar hair. 
We will also discuss globally regular solutions that exist on $x\in [0:\infty)$. Without and with backreaction, these are charged $Q$-balls and charged boson stars, respectively.

\subsection{Black hole solutions}
Here, we will discuss space-times that contain a horizon. We begin by discussing $Q$-clouds on Schwarzschild black holes
and will demonstrate that the results obtained in \cite{brihaye_hartmann2020} are not specific to the form of the scalar potential 
and that the existence of two branches of $Q$-cloud solutions can also be observed when considering a potenial of the form
(\ref{eq:potential}). We will then discuss the backreaction of these two different branches of solutions onto the space-time.

\subsubsection{Solutions without scalar fields}
Without scalar fields $\psi\equiv 0$, the equations of motion have well-known black hole solutions. These are the uncharged
Schwarzschild ($Q=0$) and the charged Reissner-Nordstr\"om solution, respectively~:
\begin{equation} 
N(x)=1-\frac{2M}{x} - \frac{\alpha Q^2}{x^2} \ \ , \ \ \sigma\equiv 1 \ \ , \ \   V(x) = \frac{Q}{x_h} - \frac{Q}{x}  \ \ .
\end{equation}
The horizon(s) of this space-time are $x_{\pm} = M \pm \sqrt{M^2 - \alpha Q^2}$. For the Reissner-Nordstr\"om solution, there exists a so-called
{\it extremal limit}, the limit of maximal possible charge for the black hole which is $x_+=x_-=M = \sqrt{\alpha} Q$. 
Note that we have chosen $V(x_h)=0$ here, such that $V_{\infty}=Q/x_h$.

\subsubsection{$Q$-clouds on Schwarzschild black holes}
In this section, we consider the equations of the matter fields 
 in the background of a Schwarzschild black hole, i.e. we set $\alpha=0$ and let $\sigma\equiv 1$, $N=1-x_h/x$, where $x_h$ is the
 event horizon radius. 
In order to ensure regularity of the matter fields on the horizon, we need to impose~:
\begin{equation}
   \psi'(x_h) = \frac{x_h}{2} \frac{d U}{d \psi}(\psi(x_h)) \ \ , \ \ V(x_h) = 0 \ \ , \ \ \psi(x \to \infty) = 0 \ \ , \ \ V(x \to \infty) = V_{\infty} \ .
\end{equation} 
Note that $V(x_h)=0$ is a gauge choice implying $\omega=0$ and hence $\Omega=e V_{\infty}$. This is the so-called synchronization condition
necessary for the scalar field to be non-trivial in the black hole background.

\begin{figure}[ht!]
\begin{center}
\includegraphics[width=8.5cm]{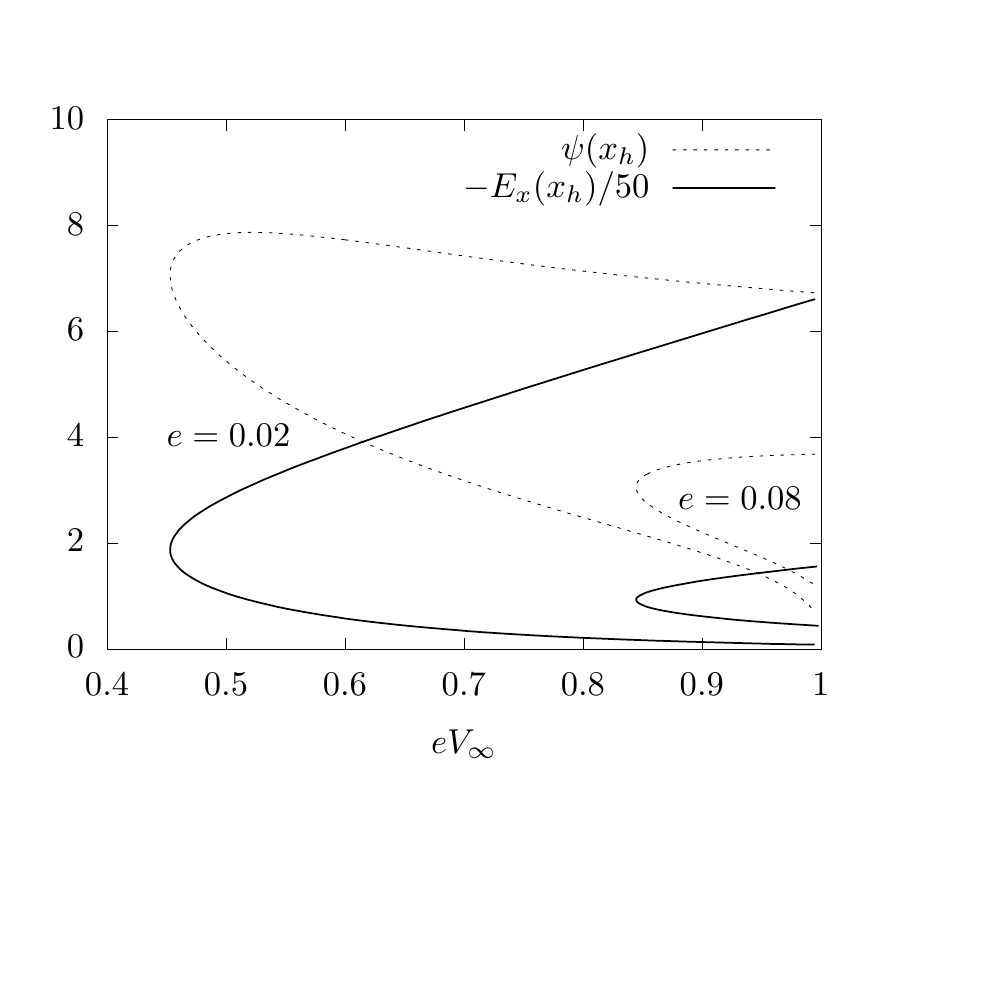}
\includegraphics[width=8.5cm]{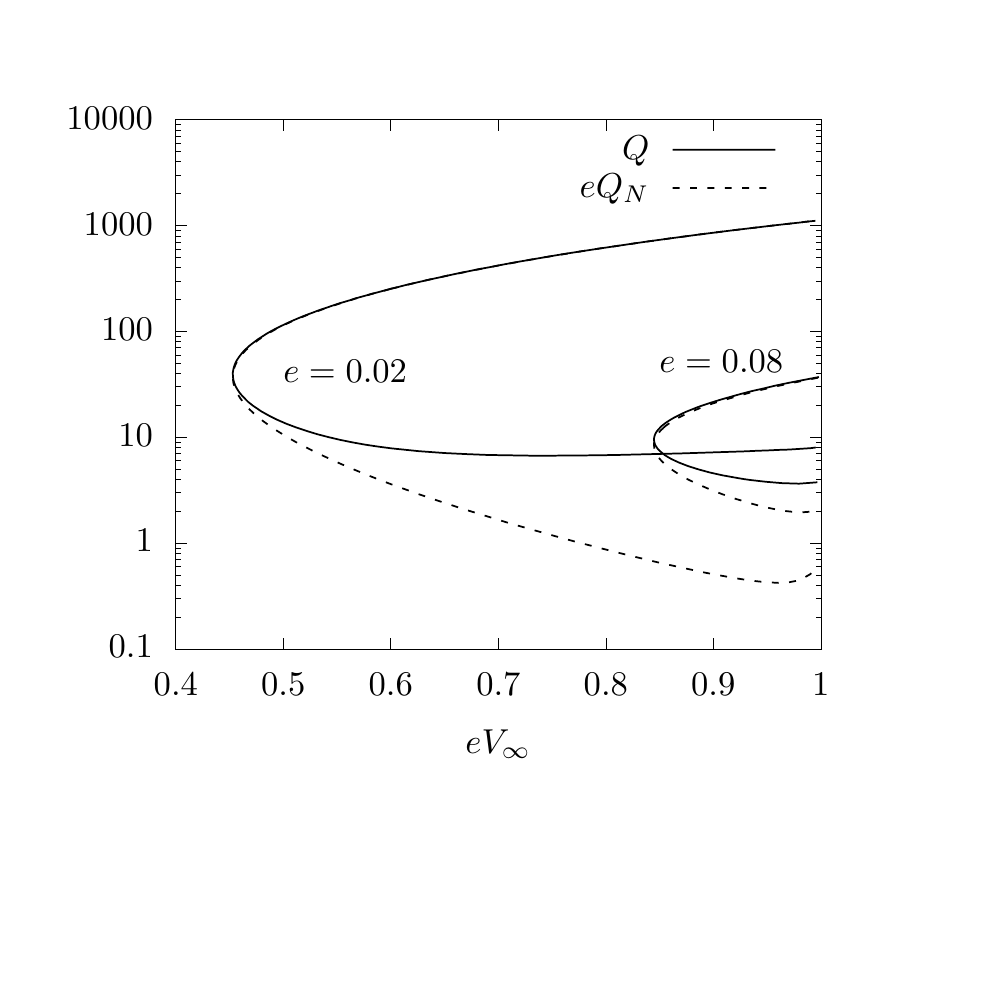}
\end{center}
\vspace{-2cm}
\caption{{\it Left}: We show the value of the scalar field $\psi$ (dashed) and the radial electric field $E_x$ (solid) on the horizon in dependence of $eV_{\infty}\equiv \Omega$ for
$x_h=0.15$ and two different values of $e$. {\it Right:} We show the values of the electric charge $Q$ (solid) and the charge contained in $Q_N$ charges $e$, $eQ_N$, (dashed) in dependence of  $eV_{\infty}\equiv \Omega$ for the same solutions. }
\label{fig:BH_cloud}
\end{figure}

In Fig.\ref{fig:BH_cloud} (left), we give the value of the scalar field $\psi$ as well as the radial electric field $E_x$ on the horizon for $x_h=0.15$
and two different values of $e$ in dependence of $eV_{\infty}$. We also give the electic charge $Q$ as well as the electric
charge contained in $Q_N$ charges $e$ in this figure (right). 
This demonstrates that charged $Q$-clouds exist on a finite interval of $\Omega\in [\Omega_{\rm\min}:1]$ and that for each choice of
$eV_{\infty}$ two solutions with different mass $M$, electric charge $Q$ and Noether charge $Q_N$ exist.
On the first branch of solutions, starting at $eV_{\infty}=1$ (which is the thresfold for the electric potential to be able to create scalar particles of mass unity) and decreasing $eV_{\infty}$, the scalar field on the horizon $\psi(x_h)$ increases up to a maximal value. With it, the radial electric field on the horizon, $E_x(x_h)$, decreases in absolute value.  Moreover, the electric charge $Q$ and the Noether charge $Q_N$ increase. 
$eV_{\infty}$ can be decreased to a minimal value that depends on $e$ and increases with the increase of $e$. Increasing $eV_{\infty}$ again from this
minimal value, a second branch of charged $Q$-clouds exist that extends all the way back to $eV_{\infty}=1$. On this second branch, the scalar field on the horizon decreases and with it the absolute value of the electric field on the horizon when increasing $eV_{\infty}$. Interesting, we observe that on this second
branch of solutions nearly all electric charge seems to be contained in the $Q_N$ scalar bosons which each carry charge $e$, while the horizon electric charge decreases indicated by the decreasing electric field on the horizon. Hence moving along the branches stores increasingly electric charge in
the scalar cloud and moves it away from the black hole horizon. Fig. \ref{fig:BH_cloud} (right) further demonstrates that
$eQ_N$ is approximately equal to $Q$ on the second branch, while on the first branch we find $e Q_N < Q$.

\begin{figure}[ht!]
\begin{center}
\includegraphics[width=8.5cm]{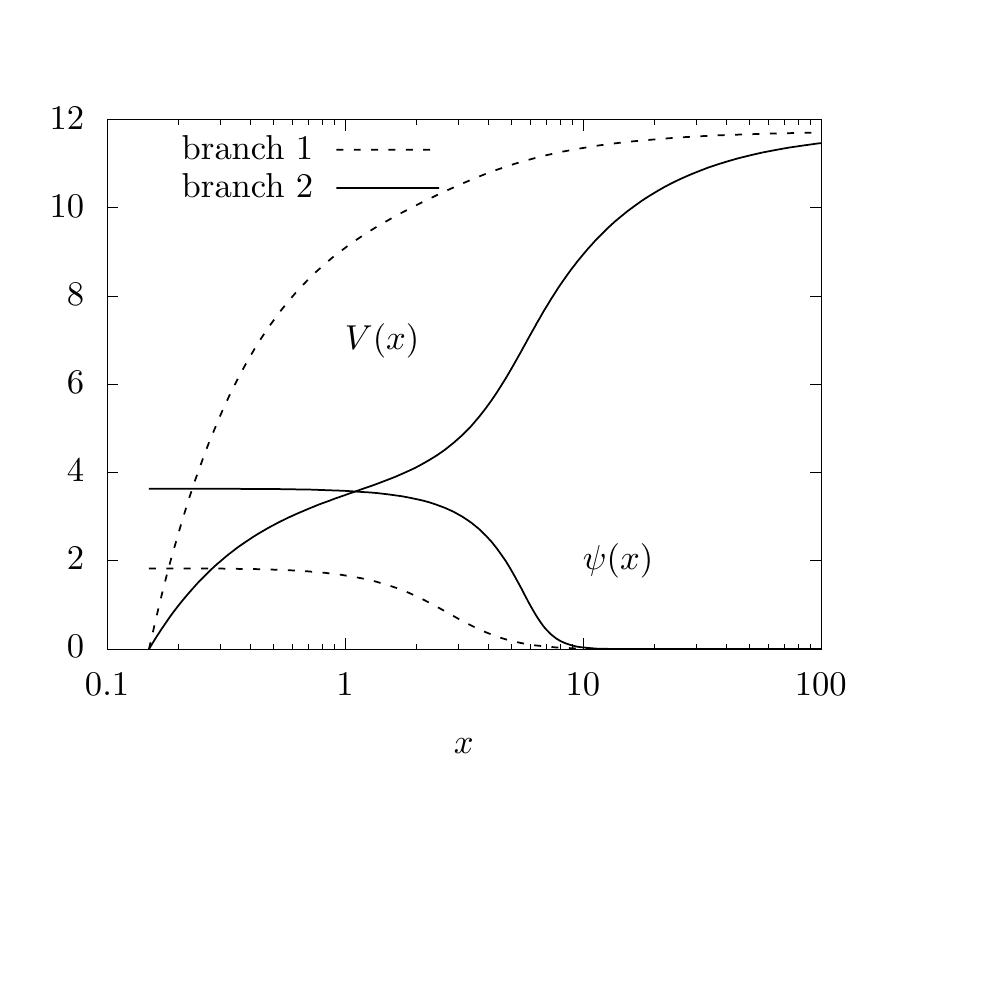}
\includegraphics[width=8.5cm]{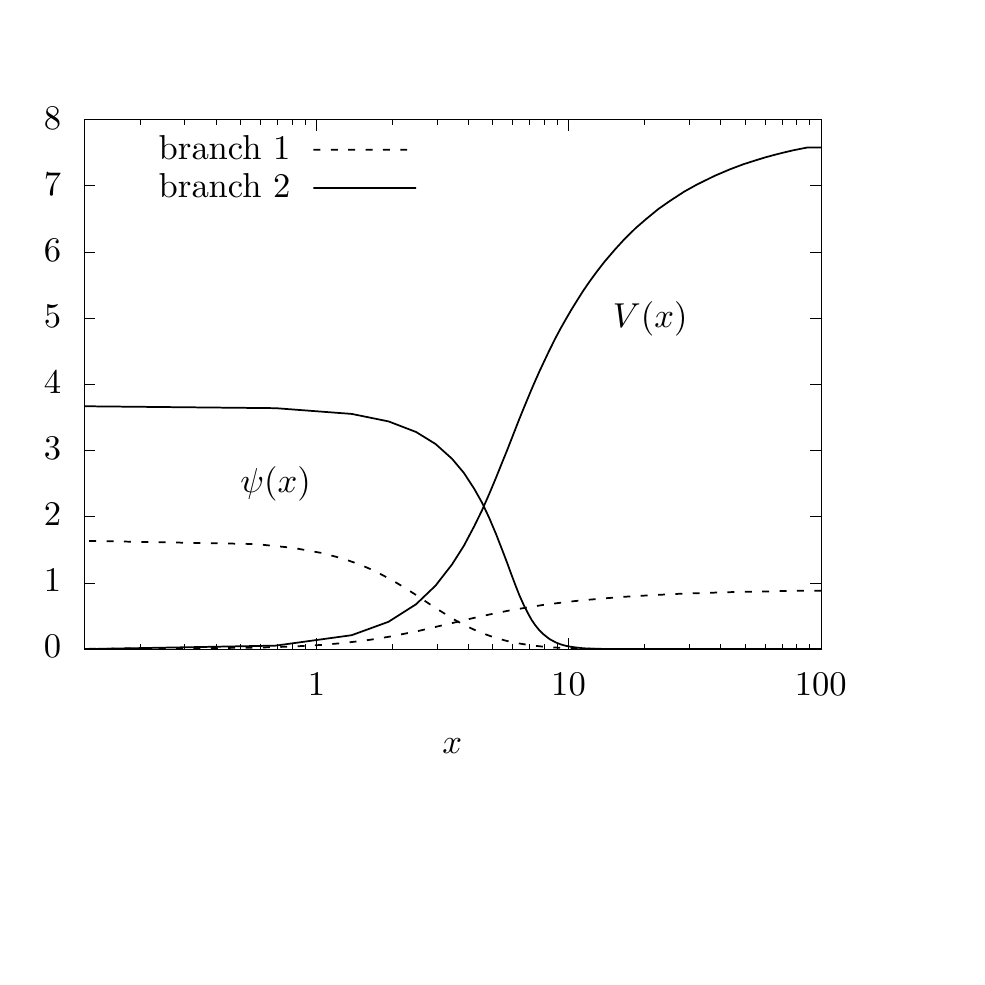}
\end{center}
\vspace{-2cm}
\caption{{\it Left}: Profiles of the scalar field $\psi(x)$ and the gauge potential $V(x)$ for the two possible $Q$-cloud solutions
on Schwarzschild black holes with $x_h=0.15$, $e=0.08$ and $e V_{\infty}=\Omega=0.94$. 
{\it Right:} Profiles of the scalar field $\psi(x)$ and the gauge potential $V(x)$ for the two possible $Q$-ball solutions
for $e=0.08$ and $\Omega=0.94$. }
\label{fig:cloud_qball_e_0_08_Omega0_94}
\end{figure}

In Fig. \ref{fig:cloud_qball_e_0_08_Omega0_94} (left) we show the profiles of the scalar field function $\psi(x)$ and the gauge field function $V(x)$ 
for the two $Q$-cloud solutions available for $e=0.08$, $eV_{\infty}\equiv \Omega=0.94$ and $x_h=0.15$. 

When fixing $V_{\infty}$ and varying $e$, we find that -- again -- two branches of solutions exist. These two branches
join at a minimal value of the gauge coupling, $e_{\rm min}$, and exist both up to $e_{\rm max}=1/V_{\infty}$ where
the effective mass $\mu_{\rm eff,\infty}$ of the scalar field becomes zero.  The minimal value of $e$ has to be determined
numerically and we find that $e_{\rm min}\approx 0.088$ for $V_{\infty}=10$ and $e_{\rm min}\approx 0.109$ for $V_{\infty}=8.8$, respectively.
In other words: the potential difference between the horizon and infinity, i.e. the chemical potential, needs to be large enough to support the scalar cloud.

\begin{figure}[h!]
\begin{center}
\includegraphics[width=10cm]{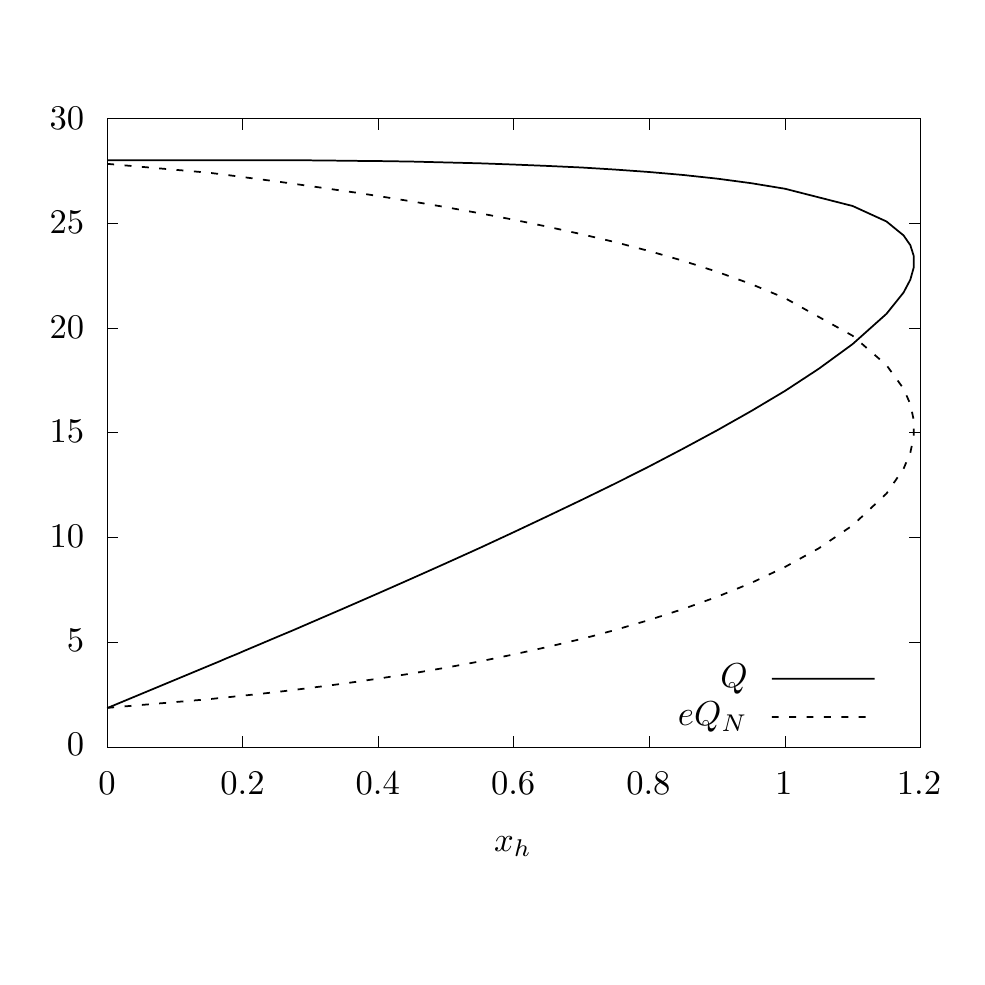}
\vspace{-1cm}
\caption{We show the electric charge $Q$ and the electric charge contained in $Q_N$ charges $e$ of $Q$-clouds on Schwarschild black holes
with $x_h=0.15$ and for $e=0.08$, $V_{\infty}=11.75$, i.e. $\Omega=eV_{\infty}=0.94$.}
\label{fig:q_qn_xh}
\end{center}
\end{figure}

Another interesting question is to understand how the surface area, i.e. the entropy of the background Schwarzschild black hole influences
the observations we have made. We have henced fixed $e$ and $V_{\infty}$ and studied the solutions for varying event horizon radius $x_h$. 
Our results are shown in Fig. \ref{fig:q_qn_xh}, where we give the electric charge $Q$ as well as the electric charge contained in the  $Q_N$ scalar bosons
that make up the cloud in dependence of the event horizon radius $x_h$. Again,we obtain two branches of solutions.  We find that the black hole needs to be sufficiently small in order to allow for the scalar clouds to exist. For the particular choice of parameters here, we find that the maximal
possible event horizon radius is $x_h\approx 1.2$. Remembering the rescalings (\ref{eq:rescalings}) this means that $r_h  < 1.2 \lambda_{{\rm C},\mu}$, where
$\lambda_{{\rm C},\mu}=1/\mu$ is the Compton wavelength of the bare scalar field. This implies that we would need an {\it ultra-light scalar
field} in order for the phenomenon of $Q$-cloud formation to be relevant for astrophysical black holes. E.g. for $\mu=10^{-10}$ eV the maximal possible
event horizon radius for $Q$-clouds to exist would be $\approx 2.4$ km, the approximate size of a solar mass black hole.
We find that when varying $eV_{\infty}$ that the maximal radius can double or triple, but it stays of the same order of magnitude. 

As a final remark, let us mention that in the limit $x_h\rightarrow 0$ we find that $Q\rightarrow eQ_N$. This is not surprising since globally
regular counterparts to the solutions discussed above exist. These are charged $Q$-balls (without backreaction of the space-time) and
charged boson stars (with backreaction), respectively. Moreover, Fig. \ref{fig:q_qn_xh} indicates that also in the regular limit
two different type of solutions should exist when fixing $e$ and $V_{\infty}$. This is what we will discuss in \ref{subsec:globally_regular}.

\subsubsection{Backreaction of $Q$-clouds}
\begin{figure}[ht!]
\begin{center}
\includegraphics[width=8cm]{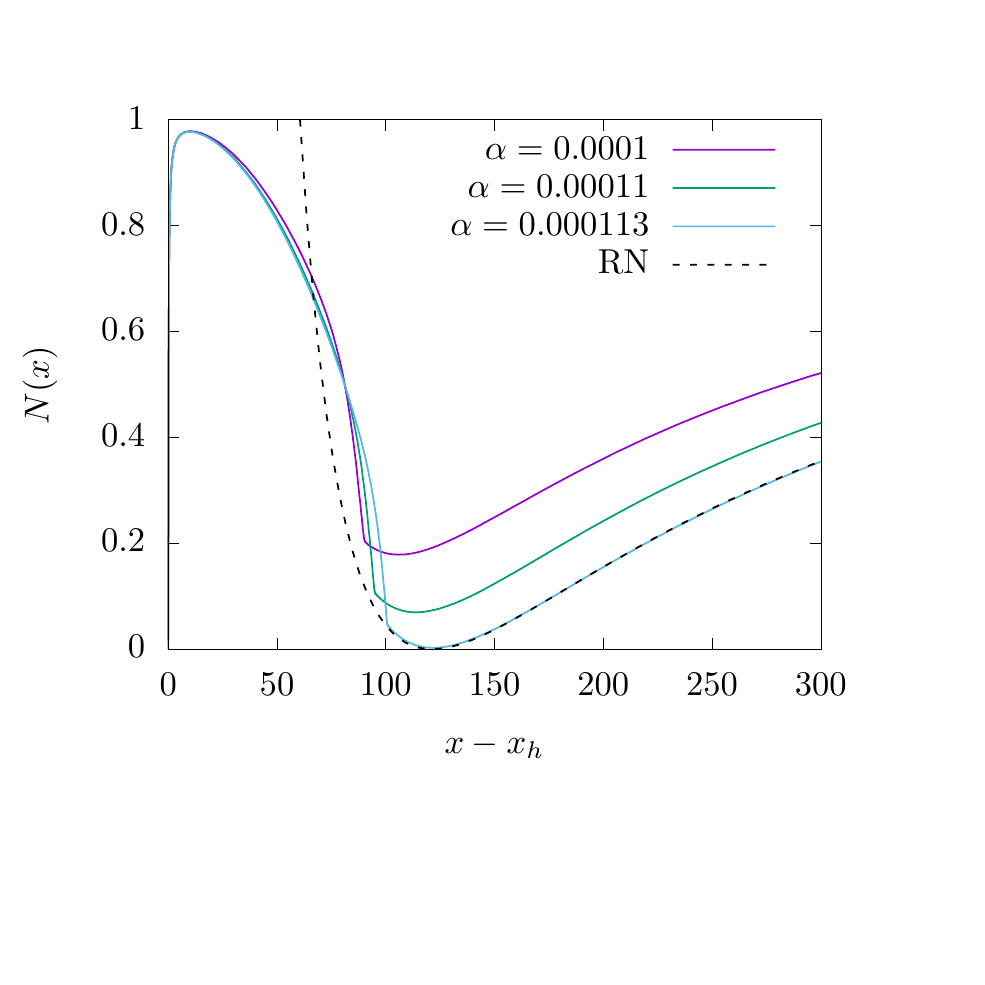}
\includegraphics[width=8cm]{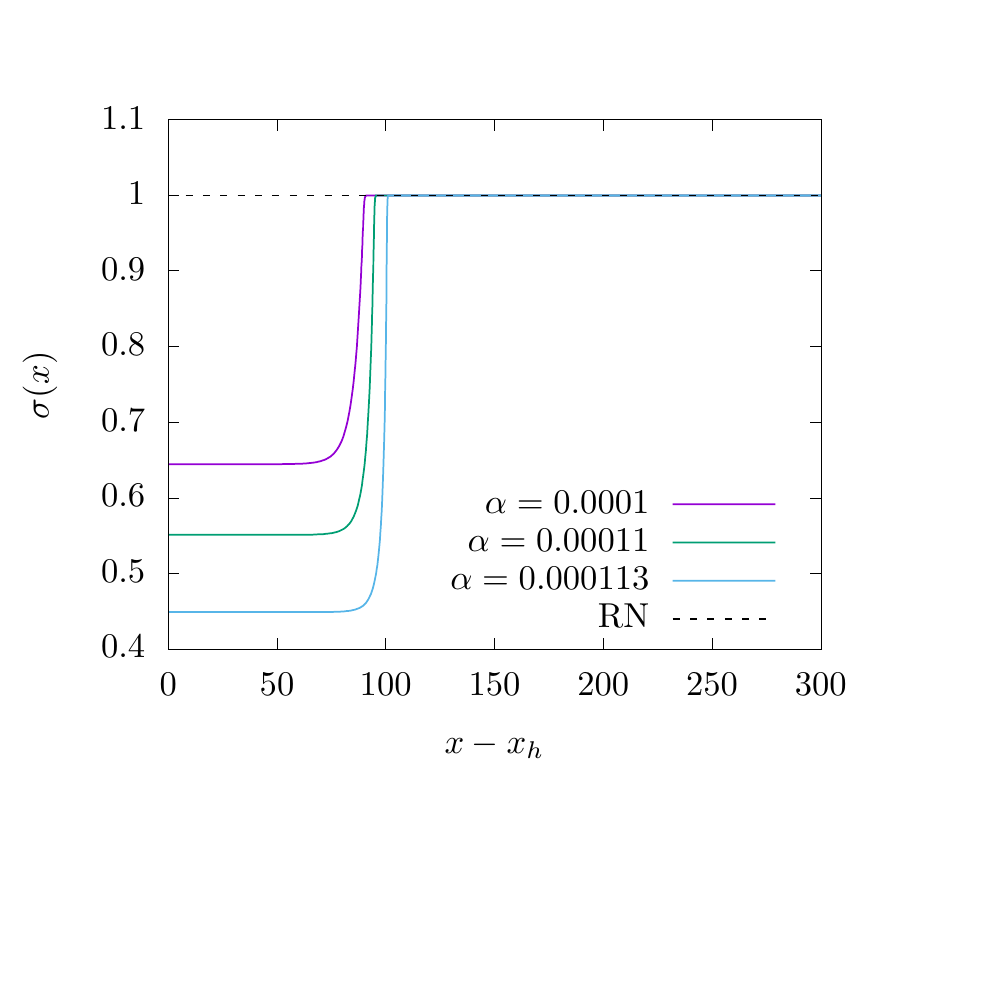}\\
\vspace{-2cm}
\includegraphics[width=8cm]{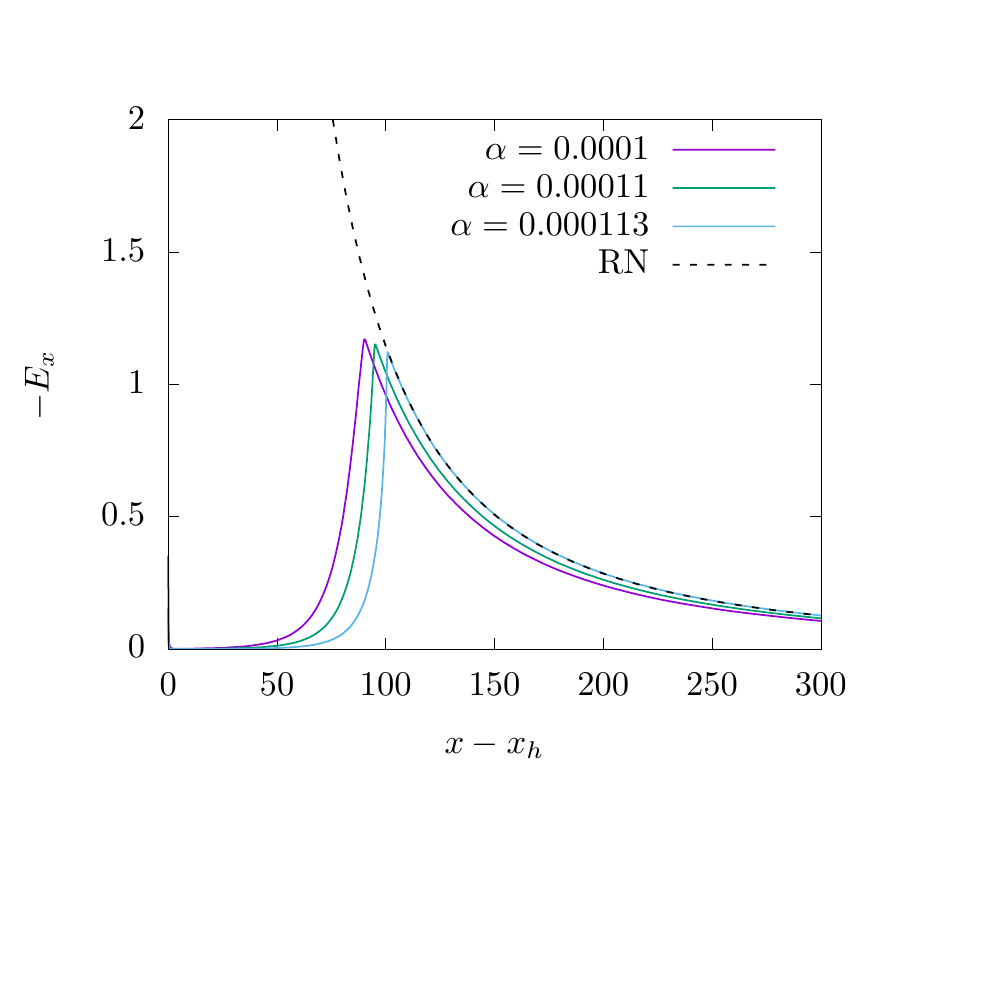}
\includegraphics[width=8cm]{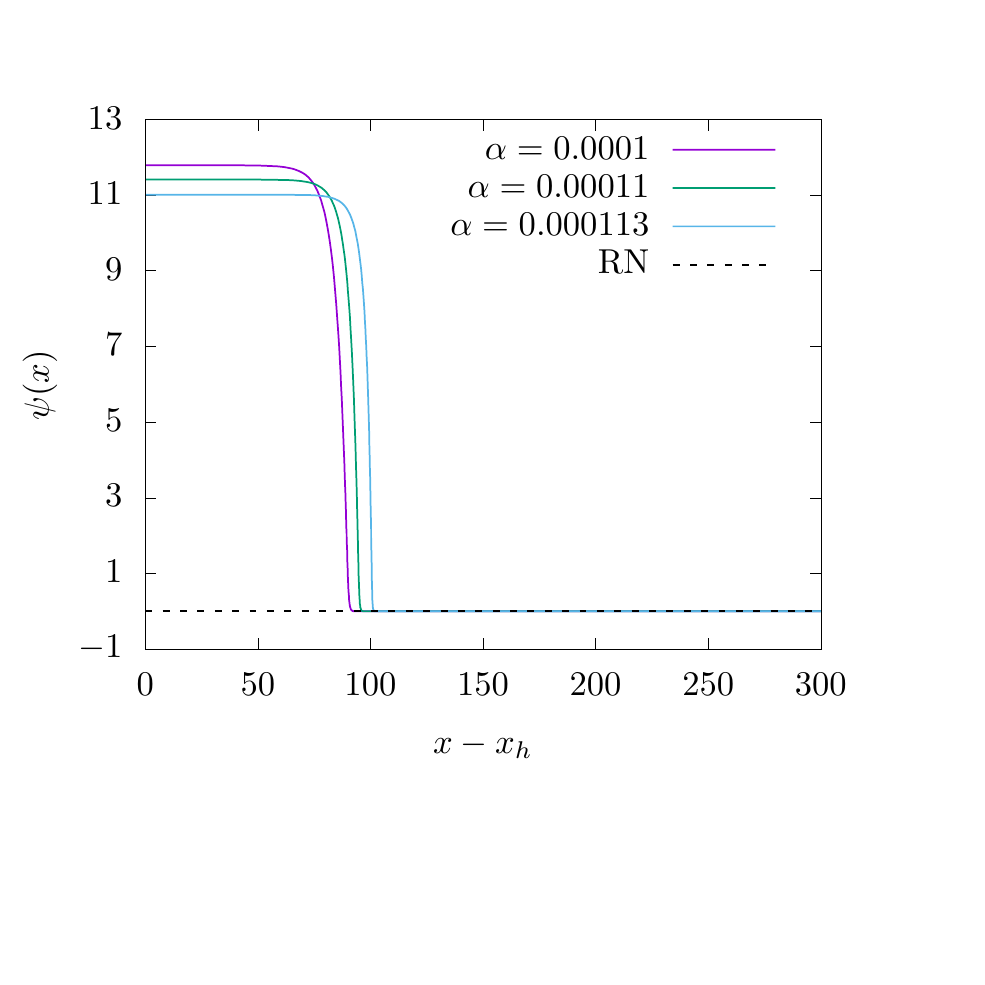}
\end{center}
\vspace{-2cm}

\caption{We show the approach to critically for a back hole with charged scalar hair for $\Omega=0.6$ and $x_h=0.15$.  }
\label{fig:BH_critical}
\end{figure}

For $\alpha > 0$, the $Q$-clouds backreact on the space-time and modify the original Schwarzschild black hole. The result is
a charged black hole which carries scalar hair on its horizon. Note that the existence of these solutions does not contradict no-hair theorems
due to the specific form of the scalar field potential.  
In \cite{brihaye_hartmann2020}  the same model as  in this paper has been discussed, but for a $\psi^6$-self-interaction potential.
It was noticed that when approaching the critical value of the gravitational coupling for solutions on the second branch of solutions (those with larger values of $\psi(x)$) the black holes form a second horizon. This second horizon can be interpreted as the horizon of an extremal
RN solution. Here, we find that this is also true when considering an exponential-type potential for the scalar field.
We show the approach to criticality for the metric and matter field functions in Fig. \ref{fig:BH_critical} for $\Omega=0.6$ and $x_h=0.15$.
We find that at a critical value of $\alpha=\alpha_{\rm cr}$, the solution forms a second horizon at $x=x_h^{(\rm ex)}$ that agrees with the
value of an extremal horizon of the corresponding RN solution. Our numerical results indicate that $x_h^{\rm (ex)}\approx M \approx \sqrt{\alpha} Q\approx 121$ for our choice of couplings.

\begin{figure}[ht!]
\begin{center}
\includegraphics[width=8.5cm]{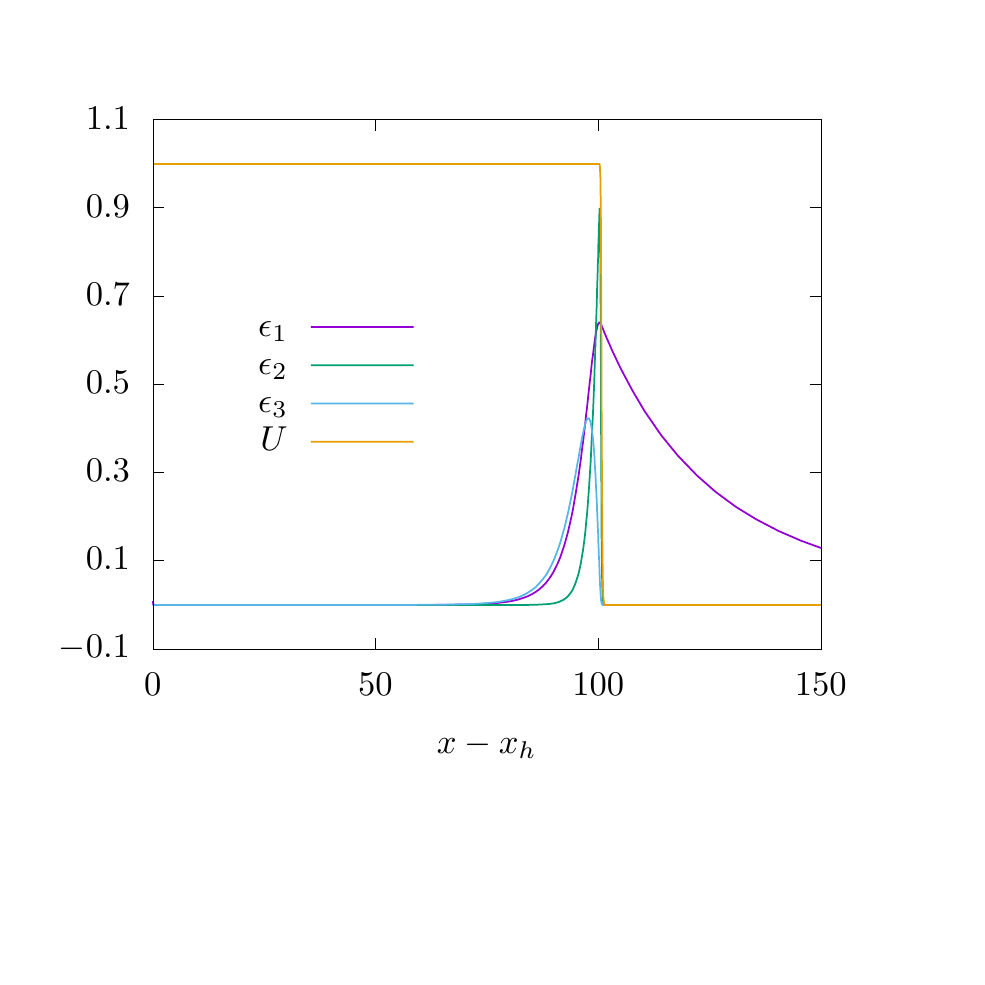}
\includegraphics[width=8.5cm]{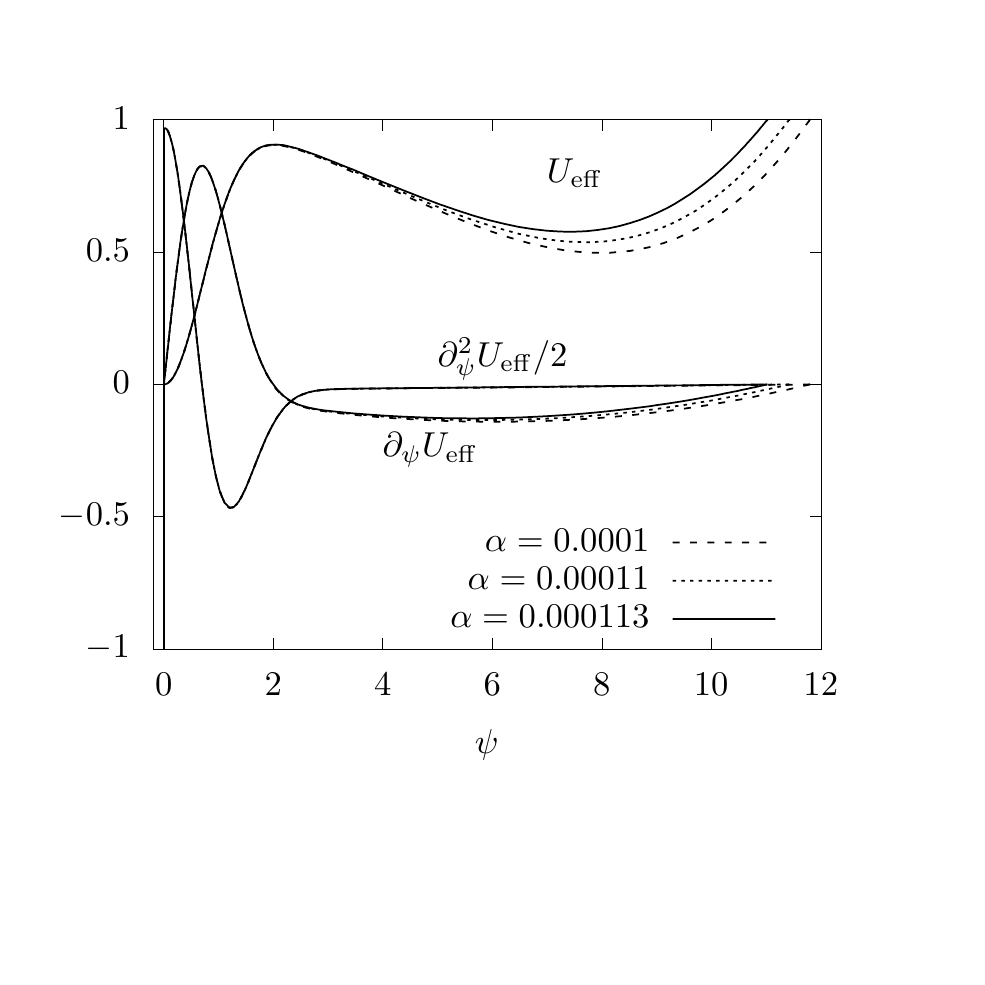}
\end{center}
\vspace{-2cm}
\caption{{\it Left:} We show the profiles of the energy density components (see  (\ref{eq:EMcomponents})) for $x_h=0.15$, $\Omega=0.6$
and close to the critical value of $\alpha$, $\alpha_{\rm cr}\approx 0.000113$. 
{\it Right}: We show the effective potential $U_{\rm eff}=U- \frac{(\omega-eV)^2}{N\sigma^2}\psi^2$ and the first ($\partial_{\psi} U_{\rm eff}$) and second derivative ($\partial^2_{\psi} U_{\rm eff}$) with respect to $\psi$
at the approach to criticality for $x_h=0.15$ and $\Omega=0.6$. }
\label{fig:ueffBH}
\end{figure}

The question then remains how these solutions can be interpreted and why they exist.
In fact, this can be understood when considering the energy density components, see (\ref{eq:EMcomponents}). In Fig. \ref{fig:ueffBH} (left),
we plot the different components very close to the critical value of $\alpha=\alpha_{\rm cr}\approx 0.000113$. 
Clearly, the scalar field energy $U(\psi)$ dominates all the other energy momentum components up to a radius $x\approx \tilde{x}$, with $\tilde{x}$ on the order
of $100$ for this particular choice of $\alpha$ and $\Omega$. This means that the energy-momentum tensor is approximately of the form $T_{\mu}^{\nu}={\rm diag}(\epsilon,p,p,p)$ where
$\epsilon=-p=U(\psi)$. This is the energy of a perfect fluid that has the equation of state of a positive cosmological constant.
In fact, assuming that $U(\psi)=U_0$ dominates the energy density up to $x=\tilde{x}$, we can integrate the equation for $m$ to get
\begin{equation}
N_{cr}(x)\approx 1- \frac{2}{3}\alpha U_0  x^2  \ ,
\end{equation}
while $\sigma=\sigma_0 \neq 0$.  This is a space-time with a positive cosmological constant $\Lambda=2\alpha U_0$. 
For our choice of potential $U_0=1$. Using $\alpha_{\rm cr}\approx 0.000113$, the cosmological horizon
of this solution would be at $x=x_c\approx 115\approx \tilde{x}$. This is not equal to the value of the extremal horizon  $x_h^{\rm (ex)}$. 
This is different to what is found in the case of inflating monopoles (see e.g. \cite{Breitenlohner:1991aa, Breitenlohner:1994di}) or
BH solutions with scalar hair in a scalar-tensor gravity model (see \cite{jutta_et_al_2020} in this issue), where $x_c=x_h^{\rm (ex)}$. 
We rather find a ``transition zone'' between $\tilde{x}$ and the extremal RN solution that is of finite thickness
 -- in our dimensionless coordinates $\Delta x :=x_h^{\rm (ex)}-\tilde{x}= {\cal O}(10)$.
 
Fig. \ref{fig:ueffBH} (right) further demonstrates that the reason for the existence of these solutions is related to the coupling of the
scalar field to the electromagnetic field. The ``bare'' potential $U(\psi)$ does not possess extrema away from $\psi=0$, while the
effective potential $U_{\rm eff}=U(\psi)-\frac{(\omega-eV)^2}{N\sigma^2} \psi^2$ does. For $\alpha$ close to $\alpha_{cr}$ and $\Omega=0.6$, there exists a local minimum for $\psi= {\cal O}(10)$. This is the value that $\psi$ takes on the horizon and up to $x\approx x_c$. 
Moreover, the first and second derivative of the effective potential with respect to $\psi$, $\partial_{\psi} U_{\rm eff}$ and $\partial^2_{\psi} U_{\rm eff}$, respectively, are smaller than the potential itself close to this value of $\psi$. Hence, the effective potential is sufficiently ``flat'' to allow
for inflation. 

\subsection{Globally regular solutions}
\label{subsec:globally_regular}
For globally regular solutions, we have to impose boundary conditions at the origin $x=0$ which
ensure regularity. These are
\begin{equation}
\label{eq:bc_origin}
m(0)=0 \ \ , \ \  V(0)=0 \  \ , \ \ V'\vert_{x=0} = 0  \ \ , \ \ \psi'\vert_{x=0}=0   \ .
\end{equation}
The condition $V(0)=0$ is, in fact, a gauge choice. 
At spatial infinity, the boundary conditions are chosen such that the solution is asymptotically
flat and has finite energy. They read~:
\begin{equation}
\label{ref:bc_infinity}
\psi(x\rightarrow \infty) \rightarrow 0 \ \ , \ \   N(x\rightarrow \infty)\rightarrow 1 \ .
\end{equation}
Note, in particular, that now $\omega\neq 0$ is necessary for globally regular solutions to exist and that it cannot be set to zero as in the black hole case.
In the following, however, we will only use the gauge-invariant quantity $\Omega=\omega - eV_{\infty}$ to describe the solutions. 

\subsubsection{(Un)Charged $Q$-balls}
\begin{figure}[ht!]
\begin{center}
\includegraphics[width=8.5cm]{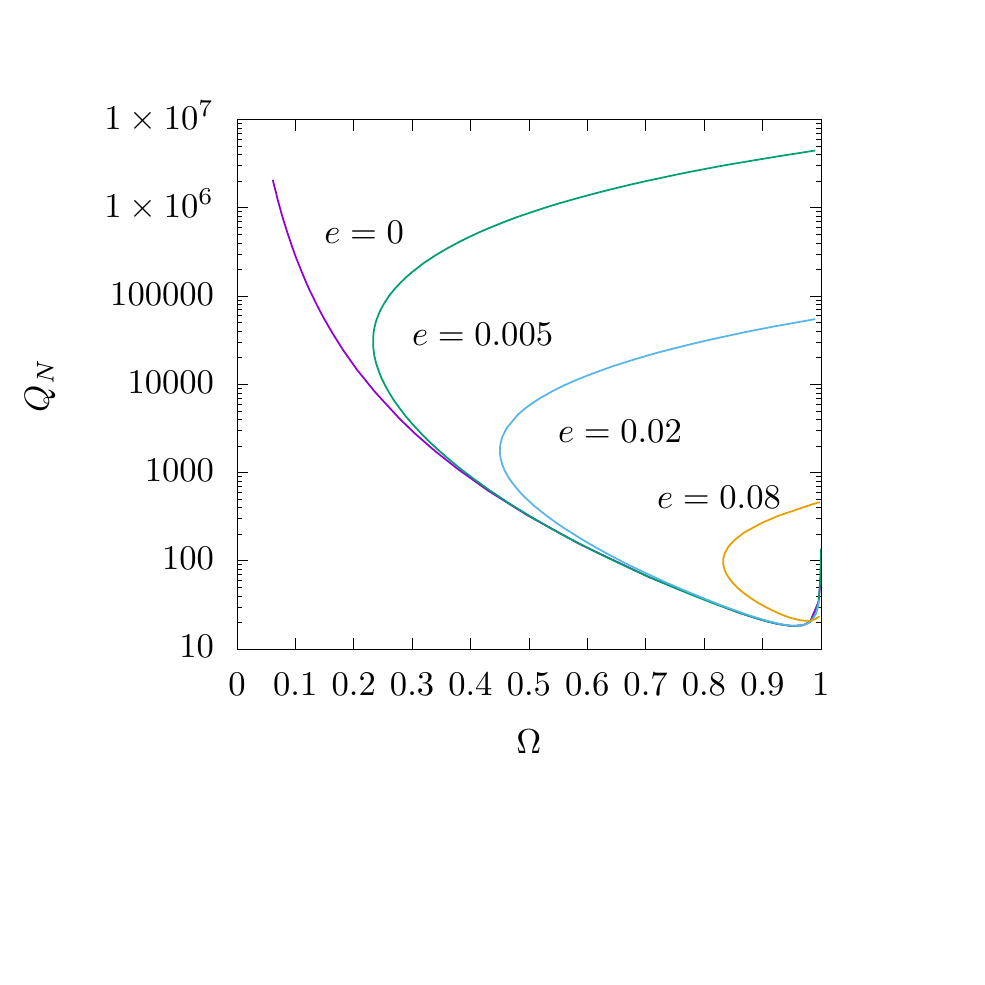}
\includegraphics[width=8.5cm]{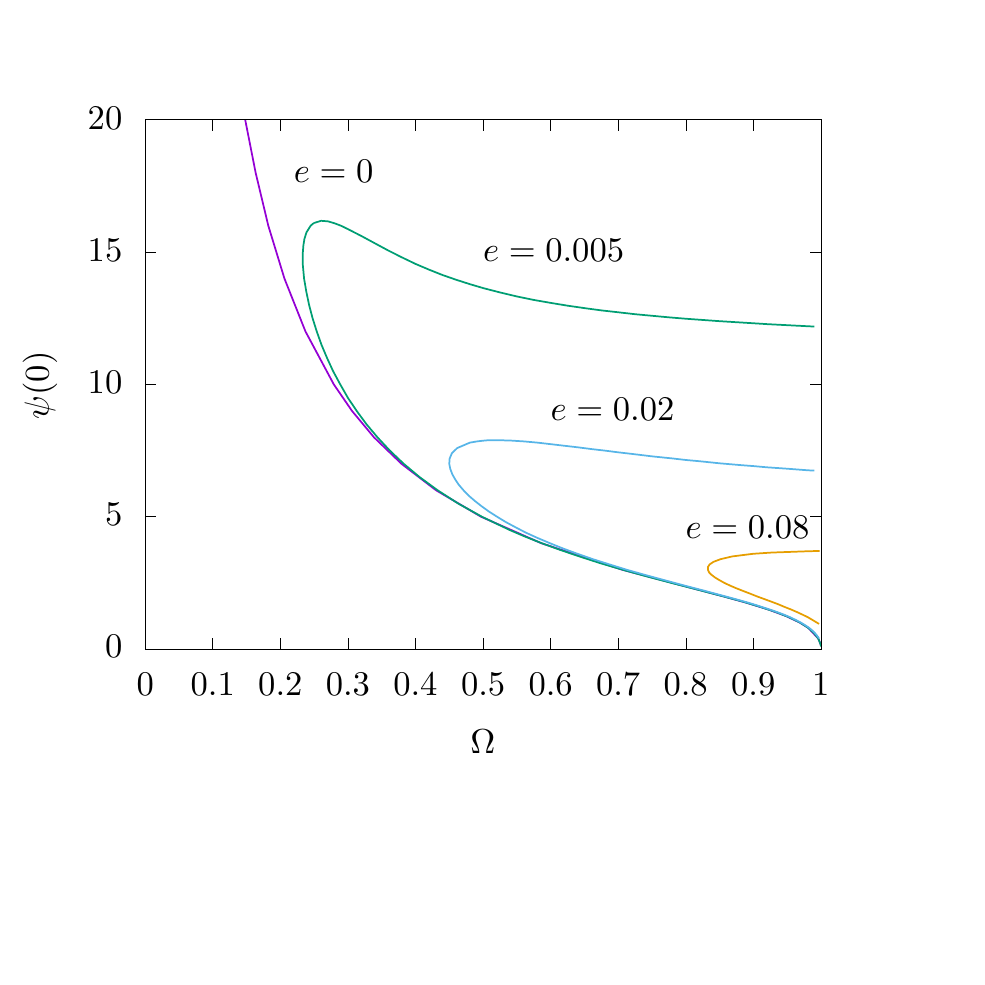}
\end{center}
\vspace{-2cm}
\caption{{\it Left}: We show the Noether charge $Q_N$ of (un)charged $Q$-balls in dependence of $\Omega$ for four different values of $e$.
{\it Right}: We show the central value of the scalar field, $\psi(0)$, of (un)charged $Q$-balls in dependence of $\Omega$ for  four different values of $e$.    }
\label{fig:qball_qn_Omega}
\end{figure}

For $\alpha=0$, the matter field equations (\ref{eq:eq_v}) and (\ref{eq:eq_psi}) decouple
from the gravity equations and the metric functions are $m\equiv 0$  (i.e. $N\equiv 1$) and $\sigma\equiv 1$ (or equal to any non-vanishing, positive constant). 
The remaining equations (\ref{eq:eq_v}), (\ref{eq:eq_psi}) possess non-trivial 
solutions, so-called {\it (charged) $Q$-balls}. 
For $V\equiv 0$ the solutions for the specific potential (\ref{eq:potential}) have been studied in 
\cite{susy1,susy2,susy3}, while for $V\neq 0$ they were investigated in \cite{Brihaye:2014gua}.
Let us remind the reader of the most important features of these solutions in the following and add details that are important for the
discussion in the following. For $e=0$, the solutions exist for
$\Omega \in (0:1]$ and  there is a one-to-one relation between the value of the
scalar field $\psi$ at the origin, $\psi(0)$, and $\Omega$. 
The limit $\psi(0) \to 0$ corresponds
to $\Omega \to 1$. In this limit, the solution becomes $\psi(x)\equiv 0$, but neither the mass nor the Noether
charge $Q_N$ of the solutions vanishes. Increasing the value $\psi(0)$ from zero, 
$\Omega$ decreases, while the mass $M$ and Noether charge $Q_N$ increase monotonically.
The dependence of $Q_N$ on $\Omega$ for $e=0$ is given in Fig. \ref{fig:qball_qn_Omega} (left). This demonstrates that for $\Omega \rightarrow 0$ the Noether
charge $Q_N$ diverges with $\psi(0)$ tending to infinity. That means that the central density of the uncharged $Q$-ball 
is not restricted and can become arbitrarily large.

When choosing $e\neq 0$ charged $Q$-balls can be constructed. However, there is a very crucial difference
to the uncharged case:  the central value of $\psi(0)$ is limited to a finite value. As such, solutions exist
only for $\Omega\in [\Omega_{\rm min}:1]$ with $\Omega_{\rm min} > 0$ for $ e\neq 0$. Moreover,  two charged $Q$-ball solutions -- and not one as in the uncharged case -- are possible. 
Fig. \ref{fig:qball_qn_Omega} demonstrates the existence of this second branch of solutions for several values of $e$. 
Here, we give the Noether charge $Q_N$ of the (un)charged $Q$-balls (including the uncharged case $e=0$) (left)
as well as the central value of the scalar field, $\psi(0)$, for several values of $e$ and in dependence of $\Omega$.
Note that this is qualitatively similar to the case for $Q$-clouds on Schwarzschild black holes, see Fig. \ref{fig:BH_cloud} (left)~: decreasing $\Omega$ on the first branch of solutions, the value of $\psi(0)$ increases up to a maximal value at $\Omega_{\rm min} > 0$. From there, a second branch
of solutions extends all the way back to $\Omega=1$. On this second branch, the value of $\psi(0)$ slightly decreases when increasing $\Omega$. 
Moreover, we find hat $\Omega_{\rm min}$ is more or less equal when considering charged $Q$-clouds and $Q$-balls, respectively.
The presence of the horizon in the former case does not seem to influence this value. The Noether charge $Q_N$ 
of the $Q$-balls on the second branch is much larger than that of solutions on the first branch. Hence, the $Q$-balls on the
second branch are composed of many more scalar bosons than those on the first branch. 

When two solutions exist for the same value of $\Omega$, we will use the convention in the following to denote  ``branch 1'' (resp. ``branch 2'')
the branch of solutions with the lower (resp. higher)  Noether charge $Q_N$.
The profiles of the scalar field $\psi(x)$ and the electric potential $V(x)$ of the two different $Q$-ball solutions  for $e=0.08$ and $\Omega=0.94$ 
are shown in Fig. \ref{fig:cloud_qball_e_0_08_Omega0_94} (right). 
Although some features appear to be similar when comparing $Q$-clouds and $Q$-balls, there is one crucial difference which is apparent from 
Fig.\ref{fig:cloud_qball_e_0_08_Omega0_94}~: the electric field. For $Q$-balls it vanishes at the origin (as it should), while
it is non-zero on the horizon of the black hole that carries the charged $Q$-cloud.

\subsubsection{Charged boson stars}
\begin{figure}[ht!]
\begin{center}
\includegraphics[width=8cm]{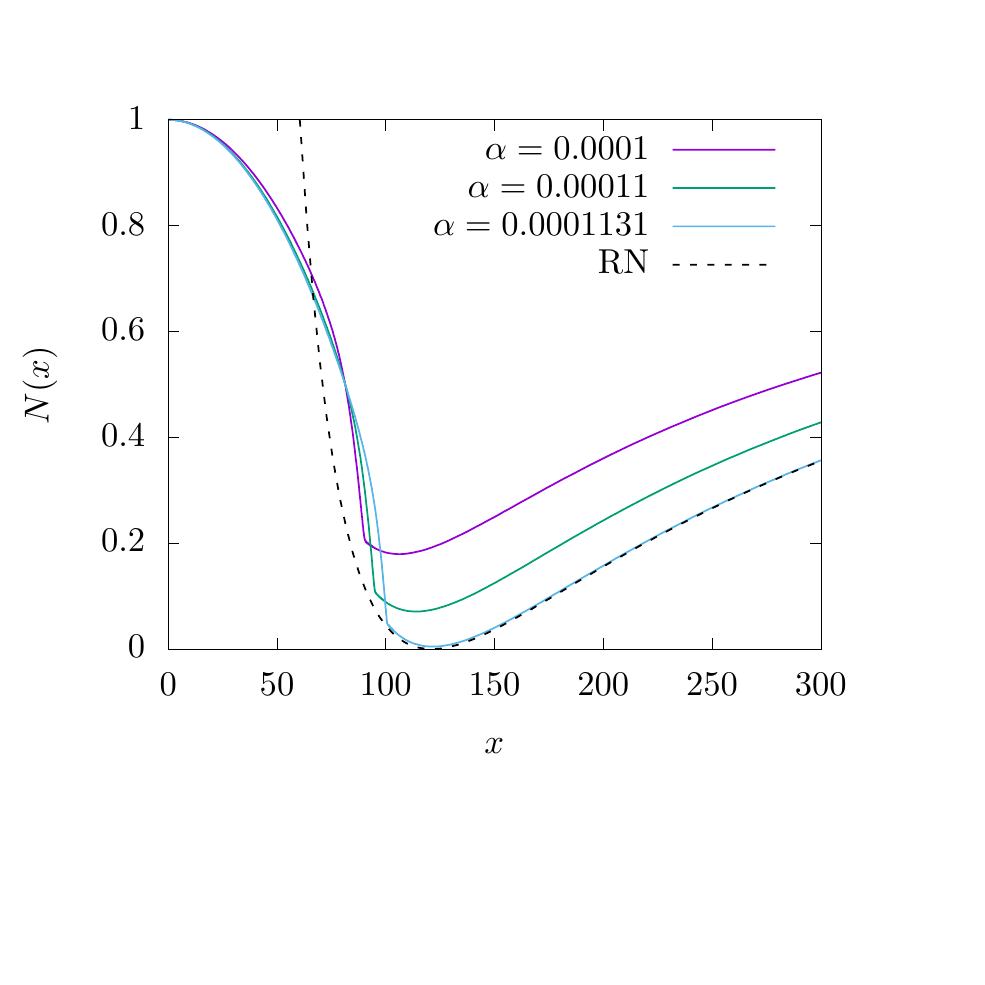}
\includegraphics[width=8cm]{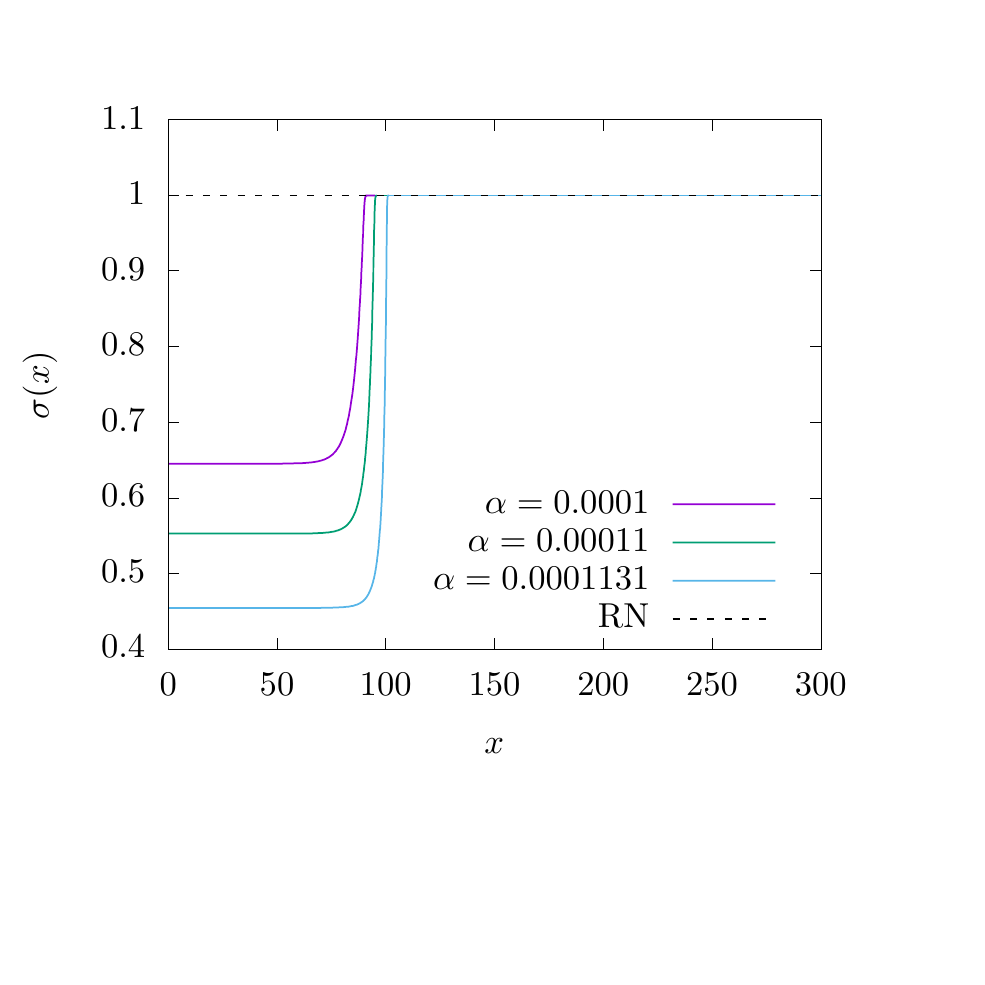}\\
\vspace{-2cm}
\includegraphics[width=8cm]{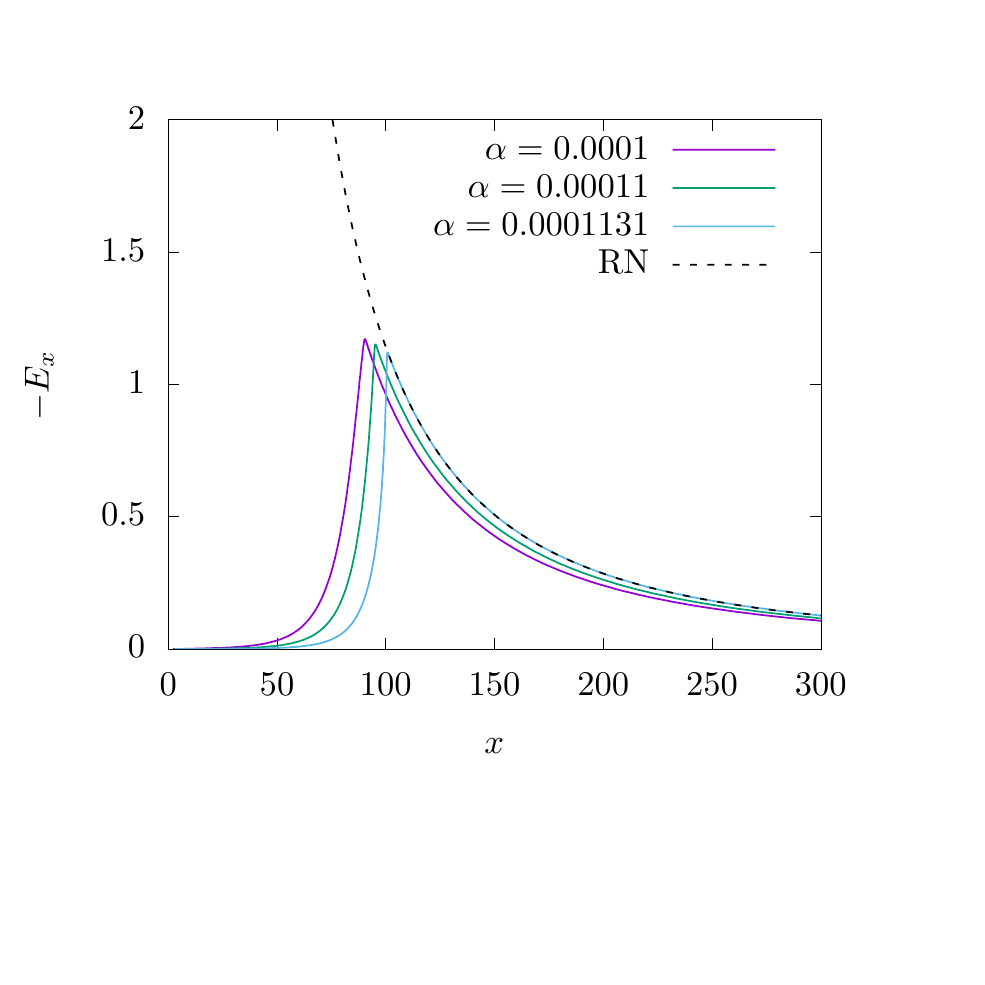}
\includegraphics[width=8cm]{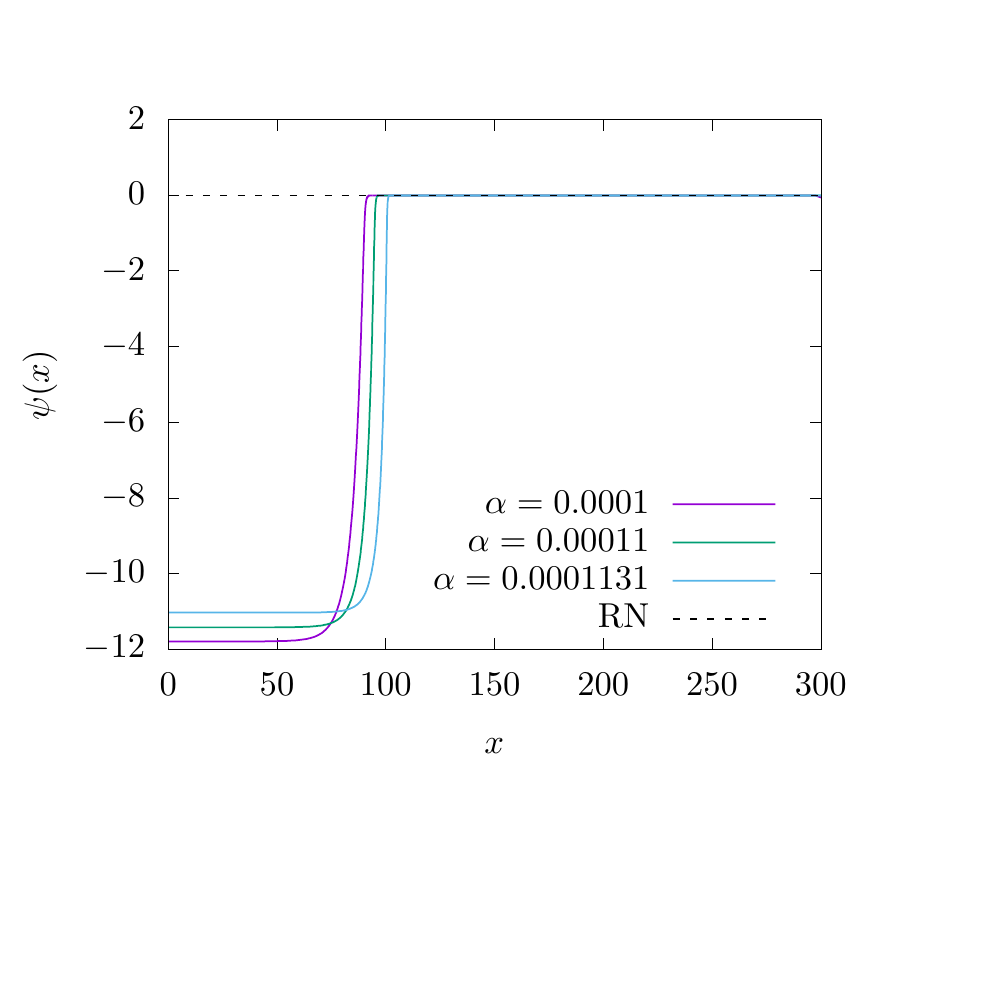}
\end{center}
\vspace{-2cm}
\caption{We show the approach to critically for a charged boson star with $\Omega=0.6$.  }
\label{fig:BS_critical}
\end{figure}

For $\alpha\neq 0$ charged boson star solutions exist. We have constructed these solutions and found that a previously unnoticed phenomenon
is present when the U(1) in the model is gauged, i.e. when the scalar field is charged under the U(1). In fact, this phenomenon is very similar to that
observed for charged black holes with scalar hair (see \cite{brihaye_hartmann2020} and discussion above). 
Choosing a solution on the second branch of solutions, where $\psi(x)$ is large at and close to the origin and increasing the gravitational
coupling leads to the appearance of a horizon at a critical value of $\alpha$. 
The approach to criticality is shown in Fig. \ref{fig:BS_critical} for $\Omega=0.6$. When increasing $\alpha$, we find that the minimal value of $N(x)$ tends
to zero, while  $\sigma(0)$ decreases (but does not reach zero). For $\alpha=\alpha_{cr}$, we find that outside a given value of the radial coordinate
$x=\tilde{x}$ the solution corresponds to the extremally charged Reissner-Nordstr\"om solution with $\sigma\equiv 1$ and $\psi\equiv 0$. For $\Omega=0.6$ we find that
$\alpha_{cr}\approx 0.000113$. Note that the value of $\tilde{x}$ and the location of the extremal
horizon at $x_h^{(ex)}\approx 121$ do not coincide in this case. In agreement with our interpretation of the limiting solution
we find that the mass at $\alpha_{cr}$ is $M=x_h^{(ex)}=\sqrt{\alpha} Q\approx 121$. Comparing this to the
case of charged black holes with scalar hair, we find that fixing $\Omega$ leads to the same numerical values of $\alpha_{\rm cr}$, $x_c$ and $x_h^{(ex)}$ for both cases. This suggestes that the phenomenon of inflation observed for black holes has a regular
limit for $x_h\rightarrow 0$. The reason for this is also connected to the observation that, while for boson stars
we always have $eQ_N=Q$, this is not true for black hole. However, for the black hole solutions on the second branch, we find
$eQ_N\approx Q$, which means that (essentially) all electric charge is carried by the scalar field. Hence, these black holes
behave very similar to charged boson stars.

The interpretation of the solutions is, hence, very similar, although we are now dealing with an a priori {\it globally regular} space-time.
In particular, for the same choice of $\Omega$ and $\alpha$, Fig. \ref{fig:ueffBH} (left and right) is identical to the corresponding equivalent plots for
the boson stars. Note that while the component $\epsilon_1$ is zero for the boson stars {\it by construction}, it is 
zero for the black hole in this particular limit. To re-emphasize this point, only in this particular limit does the
electric field on the horizon of the black hole vanish indicating that the horizon electric charge is vanishing. 

As we have discussed above, the exterior of the black hole is hence inflating with the scalar field energy dominating the remaining
energy-momentum components and hence driving inflation. We find the interior of the boson star defined by $U(\psi)=U_0\approx 1$ (as compared to the exterior of the boson star, where $U(\psi)=0$) to be inflating. Hence, {\it these solutions are the non-topological equivalent of the inflating topological defects (such as magnetic monopoles) that have been discussed in \cite{vilenkin,linde}.}  These latter defects contain
a core that is trapped inside the false vacuum of the model and this is exactly what provides the vacuum energy to drive inflation. Consequently, in these models of {\it topological inflation}, topological defects can ``blow up'' to sizes of the universe.
Here, we find that a similar phenomenon happens for localized, globally regular solutions made up out of a complex valued scalar field
coupled to an Abelian gauge field.

\section{Discussion}
In this paper, we have sudied black holes and globally regular solutions in a simple scalar field model with scalar field self-interaction of exponential
type that is motivated from models of gauge-mediated supersymmetry breaking. The crucial point for the existence of the solutions
presented here is the self-interacting of the scalar field as well as the coupling to the electromagnetic field.
This latter coupling allows for the effective scalar field potential to possess a local minimum (the ``false vacuum'') in which
the scalar field can become trapped for specific choices of the coupling constants such that the energy density
of these solutions becomes dominated by the scalar field energy of this false vacuum. When letting these solutions backreact on the space-time
we find that for sufficiently strong gravitational coupling a second horizon starts to form which corresponds to the horizon of an extremal
RN solution. 

In order to get an idea whether these results could be important at cosmological scale, let us go back to dimensionful couplings for a
moment. For $e=0.005$ (corresponding to the case $\Omega=0.6$ desribed above) we find that $\alpha_{\rm cr}\approx 0.000113$.
This corresponds to $\eta\approx 0.003 M_{\rm Pl}$, where $M_{\rm Pl}=G^{-1/2}$. Hence, the energy scale of the
scalar field would have to be on the order of the Grand Unification scale. The remaining details depend very much on the
mass $\mu$ of the scalar field. $e=0.005$ then implies $q\approx 1.67 (\mu/M_{\rm Pl})$ and a Hubble constant associated to the
expansion of $H\sim \sqrt{\frac{8\pi G U_0}{3}} =\sqrt{\frac{8\pi}{3}} \frac{\mu\eta}{M_{\rm Pl}} \approx   0.0087 \mu$. 

\vspace{2cm}

{\textbf Acknowledgments} B. H. would like to thank FAPESP for financial support under grant 2019/01511-5 as well as the
DFG Research Training Group 1620 Models of Gravity for financial support.


 \end{document}